# Fastest Distributed Consensus on Star-Mesh Hybrid Sensor Networks

Saber Jafarizadeh
School of Electrical & Information Engineering
University of Sydney
Sydney, NSW 2006, Australia
jafarizadeh@ee.usyd.edu.au

Abbas Jamalipour
School of Electrical & Information Engineering
University of Sydney
Sydney, NSW 2006, Australia
abbas@ee.usyd.edu.au

*Abstract*—Solving Fastest Distributed Consensus (FDC) averaging problem over sensor networks with different topologies has received some attention recently and one of the well known topologies in this issue is star-mesh hybrid topology. Here in this work we present analytical solution for the problem of FDC algorithm by means of stratification and semidefinite programming, for the Star-Mesh Hybrid network with K-partite core (SMHK) which has rich symmetric properties. Also the variations of asymptotic and per step convergence rate of SMHK network versus its topological parameters have been studied numerically.

## I. INTRODUCTION

A sensor network consists of a large number of sensors, called nodes, which are densely deployed inside or close to a phenomenon. The main goal of using sensor networks is to reach a global decision or estimation about the state of the phenomenon reducing the average probability of error. A limitation that has to be considered is that making global decisions (i.e., based on information collected by all sensors) has to be done through local communication between neighboring sensors, this problem is widely known as distributed detection in sensor networks. One of commonly used methods in this issue is distributed consensus averaging algorithm [1] where some of its applications include distributed agreement, synchronization problems, [2] and load balancing in parallel computers [3].

Most of the methods proposed so far deal with the FDC averaging algorithm problem by numerical convex optimization methods and to the authors' best knowledge in general no closed-form solution for finding FDC has been offered up to now, except the previous works by author and [4]. In [5], the author has proposed an analytical solution for FDC problem over symmetric and complete cored star networks.

In this work we have considered Star-Mesh Hybrid topology which has recently found applications in body area sensor networks [6]. we have solved FDC problem over Star-Mesh Hybrid network with K-partite core (SMHK) which consists of similar extended symmetric star networks where their central nodes are connected to each other in a fashion to form a complete K-partite graph (e.g. see Fig.1.). The extended symmetric star network implies a star network where the branches connected to the central node are path graphs with the same length. Also the variations of asymptotic and per step convergence rate of SMHK network versus its topological parameters have been studies numerically.

Here in this work we have managed to solve FDC problem over the SMHK network by means of stratification and semidefinite programming. Our method in this paper is based on convexity of fastest distributed consensus averaging problem, and inductive comparing of the characteristic polynomials initiated by slackness conditions in order to find the optimal weights.

The organization of the paper is as follows. Section II is an overview of the materials used in the development of the paper, including FDC algorithm, graph symmetry and SDP. Section III contains the main results of the paper where SMHK network is introduced together with the corresponding evaluated Second Largest Eigenvalue Modulus (*SLEM*) and the obtained optimal weights. Section IV is devoted to the proof of main results of the paper for SMHK network. Section V presents simulations, demonstrating the changes of asymptotic and per-step convergence rate of SMHK network versus its physical parameters and section VI concludes the paper.

## II. PRELIMINARIES

This section introduces the notation used in the paper and reviews relevant concepts from distributed consensus averaging algorithm, graph symmetry and semidefinite programming.

### A. Distributed Consensus

We consider a network $\mathcal{N}$ with the associated graph $\mathcal{G} = (\mathcal{V}, \mathcal{E})$ consisting of a set of nodes $\mathcal{V}$ and a set of edges $\mathcal{E}$ where each edge $\{i,j\} \in \mathcal{E}$ is an unordered pair of distinct nodes.

The main purpose of distributed consensus averaging is to compute the average of the initial node values $\bar{x} = (\mathbf{1}\mathbf{1}^T/n)x(0)$, through the distributed linear iterations $x(t+1) = Wx(t)$. $x(0)$ is the vector of initial node values on the network. $W$ is the weight matrix with the same sparsity pattern as the adjacency matrix of the network's associated graph and $t = 0,1,2,...$ is the discrete time index (Here **1** denotes the column vector with all coefficients one).

In [1] it has been shown that the necessary and sufficient conditions for the convergence of linear iteration mentioned above is that one is a simple eigenvalue of $W$ associated with the eigenvector **1**, and all other eigenvalues are strictly less than one in magnitude. Moreover in [1] FDC problem has been formulated as the following minimization problem

$$\min_{W} \quad \max(\lambda_2, -\lambda_n)$$
$$s.t. \quad W = W^T, W\mathbf{1} = \mathbf{1}, \forall\{i,j\} \notin \mathcal{E}: W_{ij} = 0$$

where $1 = \lambda_1 \geq \lambda_2 \geq \cdots \geq \lambda_n \geq -1$ are eigenvalues of $W$ arranged in decreasing order and $\max(\lambda_2, -\lambda_n)$ is the *Second Largest Eigenvalue Modulus* (*SLEM*) of $W$, and the main problem can be derived in the semidefinite programming form as [1]:

$$\begin{aligned} \min_W \quad & s \\ s.t. \quad & -sI \preccurlyeq W - \mathbf{1}\mathbf{1}^T/n \preccurlyeq sI, W = W^T \\ & W\mathbf{1} = \mathbf{1}, \forall \{i,j\} \notin \mathcal{E}: W_{ij} = 0 \end{aligned} \quad (1)$$

We refer to problem (1) as the Fastest Distributed Consensus (FDC) averaging problem.

### B. Symmetry of Graphs

An automorphism of a graph $\mathcal{G} = (\mathcal{V}, \mathcal{E})$ is a permutation $\sigma$ of $\mathcal{V}$ such that $\{i,j\} \in \mathcal{E}$ if and only if $\{\sigma(i), \sigma(j)\} \in \mathcal{E}$, the set of all such permutations, with composition as the group operation, is called the automorphism group of the graph and denoted by $Aut(\mathcal{G})$. For a vertex $i \in \mathcal{V}$, the set of all images $\sigma(i)$, as $\sigma$ varies through a subgroup $G \subseteq Aut(\mathcal{G})$, is called the orbit of $i$ under the action of $G$. The vertex set $\mathcal{V}$ can be written as disjoint union of distinct orbits. In [7], it has been shown that the weights on the edges within an orbit must be the same.

### C. Semidefinite Programming (SDP)

SDP is a particular type of convex optimization problem. An SDP problem requires minimizing a linear function subject to a linear matrix inequality (LMI) constraint [8]:

$$\begin{aligned} \min \quad & \rho = c^T x, \\ s.t. \quad & F(x) \geq 0 \end{aligned}$$

where $c$ is a given vector, $x^T = (x_1, \ldots, x_n)$, and $F(x) = F_0 + \sum_i x_i F_i$, for some fixed Hermitian matrices $F_i$. The inequality sign in $F(x) \geq 0$ means that $F(x)$ is positive semidefinite.

This problem is called the primal problem. Vectors $x$ whose components are the variables of the problem and satisfy the constraint $F(x) \geq 0$ are called primal feasible points, and if they satisfy $F(x) > 0$, they are called strictly feasible points. The minimal objective value $c^T x$ is by convention denoted by $\rho^*$ and is called the primal optimal value.

Due to the convexity of the set of feasible points, SDP has a nice duality structure, with the associated dual program being:

$$\begin{aligned} \max \quad & -Tr[F_0 Z] \\ s.t. \quad & Z \geq 0, Tr[F_i Z] = c_i \end{aligned}$$

Here the variable is the real symmetric (or Hermitian) positive matrix $Z$, and the data $c$, $F_i$ are the same as in the primal problem. Correspondingly, matrix $Z$ satisfying the constraints is called dual feasible (or strictly dual feasible if $Z > 0$). The maximal objective value of $-Tr[F_0 Z]$, i.e. the dual optimal value is denoted by $d^*$.

The objective value of a primal (dual) feasible point is an upper (lower) bound on $\rho^*(d^*)$. The main reason why one is interested in the dual problem is that one can prove that $d^* \leq \rho^*$, and under relatively mild assumptions, we can have $\rho^* = d^*$. If the equality holds, one can prove the following optimality condition on $x$.

A primal feasible $x$ and a dual feasible $Z$ are optimal, which is denoted by $\hat{x}$ and $\hat{Z}$, if and only if

$$F(\hat{x})\hat{Z} = \hat{Z}F(\hat{x}) = 0 \quad (2)$$

which is called the complementary slackness condition.

In one way or another, numerical methods for solving SDP problems always exploit the inequality $d \leq d^* \leq \rho^* \leq \rho$, where $d$ and $\rho$ are the objective values for any dual feasible point and primal feasible point, respectively. The difference $\rho^* - d^* = c^T x + Tr[F_0 Z] = Tr[F(x)Z] \geq 0$ is called the duality gap. If the equality $d^* = \rho^*$ holds, i.e. the optimal duality gap is zero, and then we say that strong duality holds.

## III. MAIN RESULTS

This section presents the main results of the paper. Here we introduce Star-Mesh Hybrid network with K-partite core (SMHK) with the obtained optimal weights and the corresponding *SLEM*.

A SMHK network of order $(n, m, k, L)$ consists of $n \times k$ similar extended symmetric star networks which are arranged in $n$ sets each with $k$ star networks and their central nodes are connected to each other in a fashion to form a complete K-partite graph. Every extended symmetric star network consists of $L$ path networks of length $m$ connected by a bridge to a central node. A SMHK network of order $(n = 3, m = 2, k = 2, L = 3)$ is depicted in Fig.1.

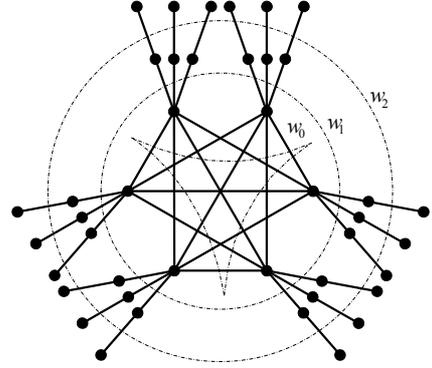

Fig.1. A SMHK network of order $(n = 3, m = 2, k = 2, L = 3)$.

In a SMHK network the optimal weight on the edges connecting central nodes of star networks together (weighted as $w_0$ in Fig.1) equals

$$w_0 = \frac{(\sqrt{n} + \sqrt{n-1}) - s(\sqrt{n} - \sqrt{n-1})}{k\sqrt{n(n-1)}(\sqrt{n} + \sqrt{n-1})}$$

and on the edges (bridges) connecting path branches to the central nodes of star networks (weighted as $w_1$ in Fig.1) the optimal weight equals

$$w_1 = \frac{\cos(\theta)(\sqrt{n} - \sqrt{n-1})\sin(m\theta) - (\sqrt{n} + \sqrt{n-1})\sin(m\theta)}{(\sqrt{n} - \sqrt{n-1})\sin((m-1)\theta) - (L+1)(\sqrt{n} + \sqrt{n-1})\sin(m\theta)}$$

where $\theta$ is the root of relation given in (\$24) and *SLEM* of network equals $\cos(\theta)$.

## IV. PROOF OF THE MAIN RESULTS

In this section solution of FDC problem and determination of optimal weights for SMHK network introduced in section IV is presented.

Here we consider a SMHK network of order $(n, m, k, L)$ with the undirected associated connectivity graph $\mathcal{G} = (\mathcal{V}, \mathcal{E})$ consisting of $|\mathcal{V}| = nk(mL + 1)$ nodes and $|\mathcal{E}| = knLm + k^2 n(n-1)/2$ edges. We denote the nodes by $(i, j, p, q)$ where $i, j, p$ and $q$ determine the set, the star network, the path branch of star network and distance from central core of the node respectively and their range are $i \in [1, n], j \in [1, k], p \in [0, L], q \in [0, m]$.

Automorphism of SMHK graph is $S_n$ permutation of sets, $S_k$ permutation of star graphs within each set and $S_L$ permutation of path graphs within each star graph hence according to subsection II-B it has $m + 1$ class of edge orbits and it suffices to consider just $m + 1$ weights $w_0, w_1, \ldots, w_m$ (as labeled in Fig.1) and consequently the weight matrix for the network can be defined as

$$W_{(i,j,p,q),(i',j',p',q')}$$
$$= \begin{cases} 1 - k(n-1)w_0 - Lw_1 & \text{for } i = i', \ j = j', \ q = q' = p = p' = 0 \\ 1 - w_q - w_{q+1} & \text{for } i = i', \ j = j', \ p = p', \ q = q' = 1, \ldots, m-1 \\ 1 - w_m & \text{for } i = i', \ j = j', \ p = p', \ q = q' = m \\ w_0 & \text{for } i \neq i', \ p = p' = q = q' = 0 \\ w_q & \text{for } i = i', \ j = j', \ p = p', \ q = 1, \ldots, m, \ q' = q - 1 \end{cases}$$

where $i, i' \in [1, n], j, j' \in [1, k]$ and $p, p' \in [0, L]$.

We associate with the node $(i, j, p, q)$ the $|\mathcal{V}| \times 1$ column vector $e_{i,j,q} = e_i \otimes e_j \otimes e_p \otimes e_q$ for $i \in [1, n], j \in [1, k], p \in [0, L]$ and $q \in [0, m]$ where $e_i, e_j, e_p$ and $e_q$ are $n \times 1, k \times 1, (L+1) \times 1$ and $(m+1) \times 1$ column vectors with one in the $i$-th, $j$-th, $(p+1)$-th and $(q+1)$-th position respectively and zero elsewhere. Introducing the new basis

$$\varphi_{\mu,\rho,\gamma,q} = \frac{1}{\sqrt{nk}} \sum_{i=1}^{n} \omega_1^{(i-1)\mu} \sum_{j=1}^{k} \omega_2^{(j-1)\rho} e_{i,j,0,q} \quad \text{for } q = 0$$

and

$$\varphi_{\mu,\rho,\gamma,q} = \frac{1}{\sqrt{nkL}} \sum_{i=1}^{n} \omega_1^{(i-1)\mu} \sum_{j=1}^{k} \omega_2^{(j-1)\rho} \sum_{p=1}^{L} \omega_3^{(j-1)\gamma} e_{i,j,p,q} \quad \text{for } q = 1, \ldots, m$$

with $\mu \in [0, n-1], \rho \in [0, k-1], \gamma \in [0, L-1]$ and $\omega_1 = e^{j\frac{2\pi}{n}}, \omega_2 = e^{j\frac{2\pi}{k}}, \omega_3 = e^{j\frac{2\pi}{L}}$ and using Stratification method [7] the weight matrix $W$ for SMHK network in the new basis takes the block diagonal form with diagonal blocks $W_1, W_2, W_3$ and $W_4$ as follows:

$$W_1 = \begin{bmatrix} 1 - Lw_1 & \sqrt{L}w_1 & 0 & & \\ \sqrt{L}w_1 & 1 - w_1 - w_2 & \ddots & & 0 \\ 0 & \ddots & & \ddots & w_m \\ & & 0 & w_m & 1 - w_m \end{bmatrix} \quad (3\text{-a})$$

$$W_2 = W_1 - k(n-1)w_0 e_1 e_1^T \quad (3\text{-b})$$

$$W_3 = W_1 - knw_0 e_1 e_1^T \quad (3\text{-c})$$

where $e_1$ is a $(m+1) \times 1$ column vector with 1 in the first position and zero elsewhere. $W_4$ obtains from removing the first row and column of $W_1$. Considering the relations between $W_1, W_2$ and $W_3$ given in (3) and the fact that $W_4$ is a submatrix of $W_1, W_2$ and $W_3$ and the *Courant-Weyl inequalities* and *Cauchy Interlacing* theorem,

*Theorem 1 (The Courant-Weyl inequalities) [9]:*

Let $A$ and $B$ be Hermitean matrices of order $n$ and let $1 \leq i, j \leq n$.
(i) If $i + j \leq n$ then $\lambda_{i+j-1}(A + B) \leq \lambda_i(A) + \lambda_j(B)$,
(ii) If $i + j - n \geq 1$ then $\lambda_i(A) + \lambda_j(B) \leq \lambda_{i+j-n}(A + B)$,
(iii) If $B$ is positive semidefinite, then $\lambda_i(A + B) \geq \lambda_i(A)$.

*Theorem 2 (Cauchy Interlacing Theorem) [10]:*

Let $A$ and $B$ be $n \times n$ and $m \times m$ matrices, where $m \leq n$, $B$ is called a compression of $A$ if there exists an orthogonal projection $P$ onto a subspace of dimension $m$ such that $PAP = B$. The Cauchy interlacing theorem states that If the eigenvalues of $A$ are $\lambda_1(A) \leq \cdots \leq \lambda_n(A)$, and those of $B$ are $\lambda_1(B) \leq \cdots \leq \lambda_m(B)$, then for all $j$, $\lambda_j(A) \leq \lambda_j(B) \leq \lambda_{n-m+j}(A)$.

we can state the following corollary for the eigenvalues of $W_1, W_2, W_3$ and $W_4$.

*Corollary 1,*

For $W_1, W_2, W_3$ and $W_4$ given in (3), theorem 1 and 2 imply the following relations between the eigenvalues of $W_1, W_2, W_3$ and $W_4$

$$\lambda_{|\mathcal{V}|}(W) = \lambda_{m+1}(W_3) \leq \lambda_{m+1}(W_2) \leq \lambda_{m+1}(W_1) \leq \lambda_m(W_4)$$

and

$$\lambda_1(W_4) \leq \lambda_1(W_3) \leq \lambda_1(W_2) \leq \lambda_1(W_1) = 1$$

It is obvious from above relations that $\lambda_2(W)$ and $\lambda_{|\mathcal{V}|}(W)$ are amongst the eigenvalues of $W_2$ and $W_3$, respectively.

In the case of $k = 1$, after stratification the weight matrix $W_2$ does not exist and consequently $\lambda_2(W)$ will not be amongst the eigenvalues of $W_3$ thus corollary 1 is true for $n, k \geq 2$ and we continue the procedure by assuming $k \geq 2$. SMHK network for $k = 1$ reduces to a complete cored star network which has been studied in [5].

Based on subsection II-A and corollary 1, one can express FDC problem for SMHK network in the form of SDP as:

$$\begin{aligned} \min \quad & s \\ \text{s.t.} \quad & W_2 \leq sI, \quad -sI \leq W_3 \end{aligned} \quad (4)$$

The matrices $W_2$ and $W_3$ can be written as

$$W_2 = I_{m+1} - k(n-1)w_0 \boldsymbol{\alpha}_0 \boldsymbol{\alpha}_0^T - \sum_{i=1}^{m} w_i \boldsymbol{\alpha}_i \boldsymbol{\alpha}_i^T \quad (5\text{-a})$$

$$W_3 = I_{m+1} - knw_0 \boldsymbol{\alpha}_0 \boldsymbol{\alpha}_0^T - \sum_{i=1}^{m} w_i \boldsymbol{\alpha}_i \boldsymbol{\alpha}_i^T \quad (5\text{-b})$$

where $\boldsymbol{\alpha}_i$ are $(m+1) \times 1$ column vectors defined as:

$$\boldsymbol{\alpha}_0(j) = \begin{cases} -1 & j = 1 \\ 0 & \text{Otherwise} \end{cases}, \quad \boldsymbol{\alpha}_1(j) = \begin{cases} \sqrt{L} & j = i \\ -1 & j = i+1 \\ 0 & \text{Otherwis} \end{cases}$$

$$\boldsymbol{\alpha}_i(j) = \begin{cases} 1 & j = i \\ -1 & j = i+1 \\ 0 & \text{Otherwise} \end{cases} \quad \text{for } i = 2, \ldots, m$$

In order to formulate problem (4) in the form of standard SDP described in section II-C, we define $F_i, c_i$ and $x$ as below

$$F_0 = \begin{bmatrix} -I_{m+1} & 0 \\ 0 & I_{m+1} \end{bmatrix}, \quad F_{m+2} = I_{2m+2}$$

$$F_1 = \begin{bmatrix} k(n-1)\boldsymbol{\alpha}_0 \boldsymbol{\alpha}_0^T & 0 \\ 0 & -kn\boldsymbol{\alpha}_0 \boldsymbol{\alpha}_0^T \end{bmatrix}$$

$$F_i = \begin{bmatrix} \boldsymbol{\alpha}_{i-1} \boldsymbol{\alpha}_{i-1}^T & 0 \\ 0 & -\boldsymbol{\alpha}_{i-1} \boldsymbol{\alpha}_{i-1}^T \end{bmatrix} \quad \text{for } i = 2, \ldots, m+1$$

$$c_i = 0, \ i = 1, \ldots m+1, \quad c_{m+2} = 1, \quad x^T = [w_0, w_1, \ldots, w_m, s]$$

In the dual case we choose the dual variable $Z \geq 0$ as $Z = \begin{bmatrix} z_1 \\ z_2 \end{bmatrix} \cdot [z_1^T \ z_2^T]$ where $z_1$, and $z_2$ are $(m+1) \times 1$ column vectors. It is obvious that $Z$ is positive definite.

From the complementary slackness condition (2) we have

$$(sI - W_2)z_1 = 0, \quad (sI + W_3)z_2 = 0 \quad (6)$$

Using the constraints $Tr[F_i Z] = c_i$, we obtain

$$k(n-1)(\boldsymbol{\alpha}_0^T z_1)^2 = kn(\boldsymbol{\alpha}_0^T z_2)^2 \quad (7\text{-a})$$

$$(\boldsymbol{\alpha}_i^T z_1)^2 = (\boldsymbol{\alpha}_i^T z_2)^2 \quad \text{for } i = 1, \ldots, m \quad (7\text{-b})$$

Considering the linear independence of $\boldsymbol{\alpha}_i$ for $i = 0, \ldots, m$, we can expand $z_1$ and $z_2$ in terms of $\boldsymbol{\alpha}_i$ as

$$z_1 = \sum_{i=0}^{m} a_i \boldsymbol{\alpha}_i, \quad z_2 = \sum_{i=0}^{m} a_i' \boldsymbol{\alpha}_i \quad (8)$$

with the coordinates $a_i$ and $a_i'$, $i = 0, \ldots, m$ to be determined.

Using (5) and the expansions (8) from comparing the coefficients of $\boldsymbol{\alpha}_i$ for $i = 0, \ldots, m$ in the slackness conditions (6), we have

$$(-s+1)a_0 = k(n-1)w_0\boldsymbol{\alpha}_0^T z_1 \quad (9\text{-a})$$

$$(-s+1)a_i = w_i\boldsymbol{\alpha}_i^T z_1, \quad (9\text{-b})$$

$$(s+1)a_0' = knw_0\boldsymbol{\alpha}_0^T z_2 \quad (9\text{-c})$$

$$(s+1)a_i' = w_i\boldsymbol{\alpha}_i^T z_2, \quad (9\text{-d})$$

where (9-b) and (9-d) holds for $i = 1, \ldots, m$. Considering (7), we obtain

$$\frac{((-s+1)a_0)^2}{k(n-1)} = \frac{((s+1)a_0')^2}{kn}$$

and $(-s+1)^2 a_i^2 = (s+1)^2 a_i'^2$ which is true for $i = 1, \ldots, m$ or

$$(a_i/a_j)^2 = (a_i'/a_j')^2 \quad (10\text{-a})$$

$$\frac{a_0^2}{k(n-1)a_i^2} = \frac{a_0'^2}{kn a_i'^2} \quad (10\text{-b})$$

for $\forall i, j = [1, m]$ and for $\boldsymbol{\alpha}_i^T z_1$ and $\boldsymbol{\alpha}_i^T z_2$, we have

$$\boldsymbol{\alpha}_i^T z_1 = \sum_{j=0}^{m} a_j G_{i,j}, \quad \boldsymbol{\alpha}_i^T z_2 = \sum_{j=0}^{m} a_j' G_{i,j} \quad (11)$$

where $G$ is the Gram matrix, defined as $G_{i,j} = \boldsymbol{\alpha}_i^T \boldsymbol{\alpha}_j$ or equivalently

$$G = \begin{bmatrix} 1 & -\sqrt{L} & 0 & \cdots & 0 \\ -\sqrt{L} & L+1 & -1 & 0 & \vdots \\ 0 & -1 & 2 & \ddots & 0 \\ \vdots & 0 & \ddots & \ddots & -1 \\ 0 & \cdots & 0 & -1 & 2 \end{bmatrix}$$

Substituting (11) in (9) we have

$$(-s + 1 - k(n-1)w_0)a_0 = -k(n-1)w_0 a_1 \quad (12\text{-a})$$

$$(-s + 1 - (L+1)w_1)a_1 = -w_1(\sqrt{L}a_0 + a_2) \quad (12\text{-b})$$

$$(-s + 1 - 2w_i)a_i = -w_i(a_{i-1} + a_{i+1}) \quad (12\text{-c})$$
$$\text{for } i = 2, \ldots, m-1$$

$$(-s + 1 - 2w_m)a_m = -w_m a_{m-1} \quad (12\text{-d})$$

and

$$(s + 1 - knw_0)a_0' = -knw_0 a_1' \quad (13\text{-a})$$

$$(s + 1 - (L+1)w_1)a_1' = -w_1(\sqrt{L}a_0' + a_2') \quad (13\text{-b})$$

$$(s + 1 - 2w_i)a_i' = -w_i(a_{i-1}' + a_{i+1}') \quad (13\text{-c})$$
$$\text{for } i = 2, \ldots, m-1$$

$$(s + 1 - 2w_m)a_m' = -w_m a_{m-1}' \quad (13\text{-d})$$

Now we can determine the optimal transition probabilities in an inductive manner as follows:

In the first stage, from comparing equations (12-d) and (13-d) and considering the relation (10-a), we can conclude that $(-s + 1 - 2w_m)^2 = (s + 1 - 2w_m)^2$ which results in $w_m = 1/2$. Assuming $s = \cos(\theta)$ and substituting $w_m = 1/2$ in (12-d) and (13-d), we have

$$a_{m-1} = (\sin(2\theta)/\sin(\theta))a_m,$$

$$a_{m-1}' = (\sin(2(\pi - \theta))/\sin(\pi - \theta))a_m'$$

Continuing the above procedure inductively, up to $i - 1$ stages, and assuming

$$a_j = (\sin((m-j+1)\theta)/\sin(\theta))a_m,$$

$$a_j' = (\sin((m-j+1)(\pi-\theta))/\sin(\pi-\theta))a_m'$$

for $m - i < \forall j \leq m$ at the $i$-th stage, by comparing equations (12-b) and (13-b) and considering relation (10-a) we can conclude that

$$\left((-s + 1 - 2w_i)\sin((m-i+1)\theta) + w_i\sin((m-i)\theta)\right)^2 =$$
$$\left((s + 1 - 2w_i)\sin((m-i+1)(\pi-\theta)) + w_i\sin((m-i)(\pi-\theta))\right)^2$$

which results in

$$w_i = 1/2 \quad (14)$$

and consequently

$$a_{i-1} = a_m \times \left((\sin((m-i+2)\theta))/\sin(\theta)\right) \quad (15\text{-a})$$

$$a_{i-1}' = a_m' \times (\sin((m-i+2)(\pi-\theta))/\sin(\pi-\theta)) \quad (15\text{-b})$$

where (14) and (15) hold true for $i = 2, \ldots, m$. in the $m$-th stage from equations (12-b) and (13-b) and using (10-b) and (15), we obtain

$$w_1 = \frac{s(\sqrt{n} - \sqrt{n-1})\sin(m\theta) - (\sqrt{n} + \sqrt{n-1})\sin(m\theta)}{(\sqrt{n} - \sqrt{n-1})\sin((m-1)\theta) - (L+1)(\sqrt{n} + \sqrt{n-1})\sin(m\theta)} \quad (16)$$

and by substituting $w_1$ in (12-b) and (13-b) we obtain

$$a_0 = F(\theta)a_m, \quad a_0' = F(\pi - \theta)a_m' \quad (17)$$

where

$$F(\theta) = \left(\frac{1+L}{\sqrt{L}} + \frac{\cos(\theta) - 1}{w_1}\right)\frac{\sin(m\theta)}{\sin(\theta)} - \frac{\sin((m-1)\theta)}{\sqrt{L}\sin(\theta)} \quad (18)$$

with $w_1$ given in (16).
Finally in the last stage, from equations (12-a) and (13-a) and using relations (10-b) and (17), we can conclude that

$$w_0 = \frac{(\sqrt{n} + \sqrt{n-1}) - s(\sqrt{n} - \sqrt{n-1})}{k\sqrt{n(n-1)}(\sqrt{n} + \sqrt{n-1})} \quad (19)$$

Substituting $w_0$ and $s = \cos(\theta)$ in (12-a) we can conclude that $\theta$ has to satisfy

$$(1 - \cos(\theta))F(\theta) + \left(\frac{\sqrt{n-1}}{\sqrt{n}}\right)(1 - H(n)\cos(\theta))\left(\sqrt{L}\frac{\sin(m\theta)}{\sin(\theta)} - F(\theta)\right) = 0 \quad (20)$$

where $H(n) = (\sqrt{n} - \sqrt{n-1})/(\sqrt{n} + \sqrt{n-1})$.

Also one should notice that necessary and sufficient conditions for the convergence of the weight matrix are satisfied, since all roots of $s$ which are the eigenvalues of $W$ are strictly less than one in magnitude, and one is a simple eigenvalue of $W$ associated with the eigenvector $\mathbf{1}$, to support this fact we have computed numerically the roots of equation (20) whereby considering the relation $s = \cos(\theta)$ and that all roots of (20) are simple, we can conclude that all roots of $s$ are strictly less than one in magnitude.

## V. Numerical Results

In this section we study asymptotic and per step convergence rates of SMHK network in terms of its topological parameters, namely, number of sets $(n)$, number of star graphs in each set $(k)$ and size of star networks, namely number $(L)$ and length $(m)$ of path branches.

From (16) and (18) we can see that $w_1$ and $F(\theta)$ are independent of $k$ and according to (20) we can conclude that $k$ does not make any restrictions over *SLEM* of SMHK network.
To confirm the independence of *SLEM* from number of star graphs in each set $(k)$ numerically in Fig.2. *SLEM* of SMHK network with $m = 2$ and $L = 2$ in terms of number of sets $(n)$ and number of star graphs in each set $(k)$ is depicted.

As it is obvious from Fig.2. changing the number of star graphs in each set $(k)$ does not make any serious effect on asymptotic convergence rate of SMHK network, but SLEM of SMHK network decreases as the number of sets $(n)$ increases which is due to diminution of bottleneck effect of the core.

In Fig.3. *SLEM* of SMHK network with $n = 3$ and $k = 2$ in terms of length of path branches $(m)$ and number of path branches in each star network $(L)$ is depicted. From the results in Fig.3. it can be deduced that *SLEM* of SMHK network increases by size of the star network. In Fig.4. normalized Euclidean distance of vector of node values from the stationary distribution in terms of number of iterations for the SMHK with $n = 3, m = 2, k = 2$ and two different values of $L$ namely 3 and 20 is presented.

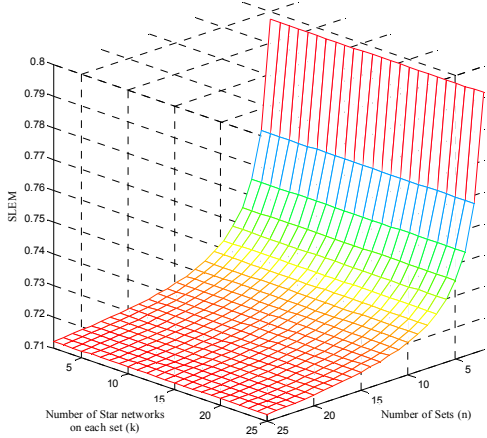

Fig.2. SLEM of SMHK network with $L = 2, m = 2$ in terms of number of sets $(n)$ and number of star graphs in each set $(k)$.

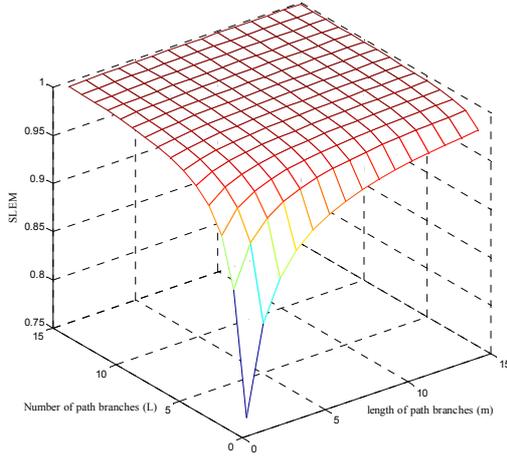

Fig.3. SLEM of SMHK network with $n = 3, k = 2$ in terms of length of path branches $(m)$ and number of path branches in each star network $(L)$.

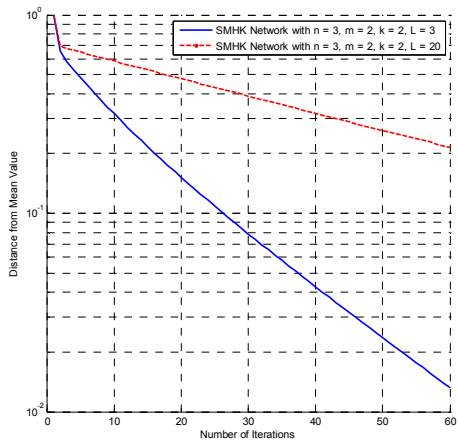

Fig.4. Normalized Euclidean Distance of vector of node values from the stationary distribution in terms of number of iterations for SMHK networks with $n = 3, m = 2, k = 2$ and $L = 3$ & $20$.

It should be mentioned that the results depicted in Fig.4. is in logarithmic scale and generated each based on 1000 trials (a different random initial node values are generated for each trial). It is obvious from Fig.4. that increasing the number of path branches of the star networks ($L$) decreases the per step convergence rate of SNHK network. Also from Fig.4. and the scale of results depicted in Fig.2. and Fig.3. it can be concluded that the changes in size of star networks can affect asymptotic and per step convergence rates of the SMHK network more than the total number of star networks.

## VI. CONCLUSION

Distributed Consensus averaging Algorithm in sensor networks has received renewed interest recently, but Most of the methods proposed so far usually avoid the direct computation of optimal weights and deal with the FDC problem by numerical convex optimization methods.

Here in this work, we have solved FDC problem for the star-mesh hybrid network with K-partite core by means of stratification and SDP. Our approach is based on fulfilling the slackness conditions, where the optimal transition probabilities are obtained by inductive comparing of the characteristic polynomials initiated by slackness conditions.

We have studied asymptotic and per step convergence rates of SMHK network in terms of its topological parameters and the simulation results confirm that changing the total number of star networks, does not make any major effects in asymptotic and per step convergence rates of SMHK network. On the contrary increasing the size of star networks namely, length and number of path branches decreases the asymptotic and per step convergence rates of the SMHK network.

We believe that the method used in this paper is powerful and lucid enough to be extended to networks with more general topologies which is the object of our future investigations.